\documentclass[11pt]{article}
\usepackage{moriond,epsfig}

\def\be{\begin{equation}}
\def\ee{\end{equation}}
\def\bea{\begin{eqnarray}}
\def\eea{\end{eqnarray}}

\newcommand{\met}{$\rlap{\kern0.25em/}E_T$ } 
\newcommand{\et}{{E$_{T}$} }
\newcommand{\pt}{{P$_{T}$} }
\newcommand{\pb}{$pb^{-1}$ }
\newcommand{\fb}{$fb^{-1}$}
\newcommand{\lumi}{$\cal{L}$ }
\newcommand{\mypm}{$\pm$ }
\newcommand{\mustar}{$\mu^*$ }
\newcommand{\ppbar}{$p\bar{p}$ }
\newcommand{\sqrts}{$\sqrt{s}$ }
\newcommand{\gev}{GeV}
\newcommand{\cgev}{GeV/c}
\newcommand{\ccgev}{GeV/c$^2$}

\begin{document}
\vspace*{4cm}
\title{NON SUSY SEARCHES AT THE TEVATRON}

\author{ PHILIPPE GRIS \\
\small{on behalf of the CDF and D0 collaborations}}

\address{Laboratoire de Physique Corpusculaire - IN2P3/CNRS, Clermont-Ferrand, France}

\maketitle\abstracts{
Recent results from searches for non-supersymmetric particles by the CDF and D0 collaborations are reported. The sample taken during the RunII of the Fermilab TeVatron collider at \sqrts=1.96 TeV is used. The integrated luminosity analyzed ranges from 300 to 1020 \pb depending on the search.}

\section{Introduction - Models and final states}

Despite its great success, the Standard Model (SM) leaves pending questions related for instance to the structure of its constituents (are leptons and quarks fundamental or composite entities?) or to fundamental interactions (what about gravitation?). Physics beyond the SM is a very broad subject with many models addressing one (sometimes more) given problem. In this paper, experimental searches inspired by non-supersymmetric models such as compositeness, technicolor or Extra Dimensions will be described. The phenomenology is quite rich, since particles such as leptoquarks, excited fermions, Z', W', Kaluza-Klein excitations or Randall-Sundrum gravitons are expected to be produced. Detailing all possible final states would be tedious. We thus choose to present the searches performed by the CDF and D0 collaborations at the Fermilab TeVatron (\ppbar collider at \sqrts=1.96 TeV) through three classes of final states: leptons/photons, leptons+jets and jets+missing energy(\met).

\section{Leptons/photons final state}

\begin{figure}[htbp]
\begin{center}
\includegraphics[width=4.cm]{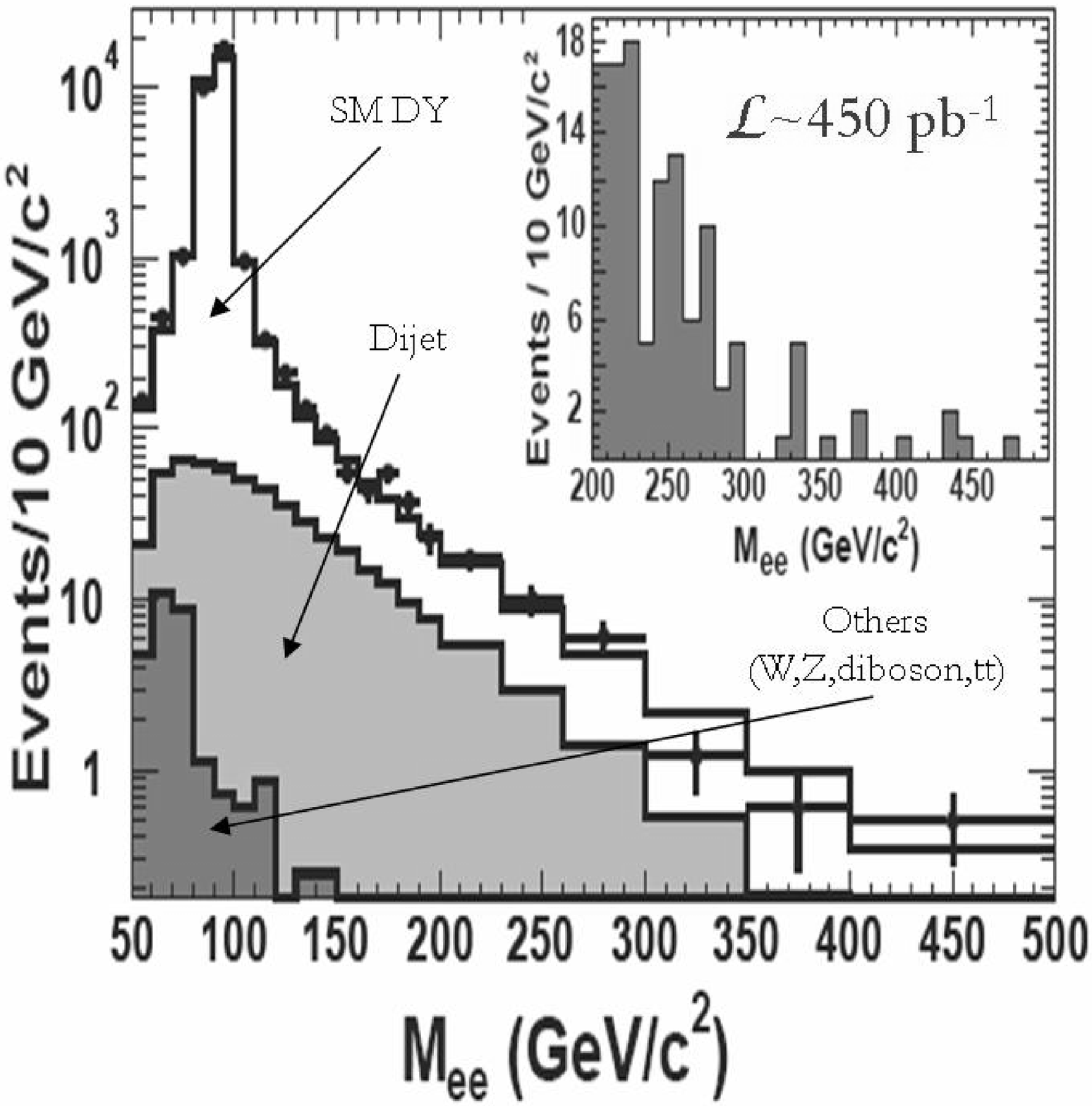}
\includegraphics[width=4.cm]{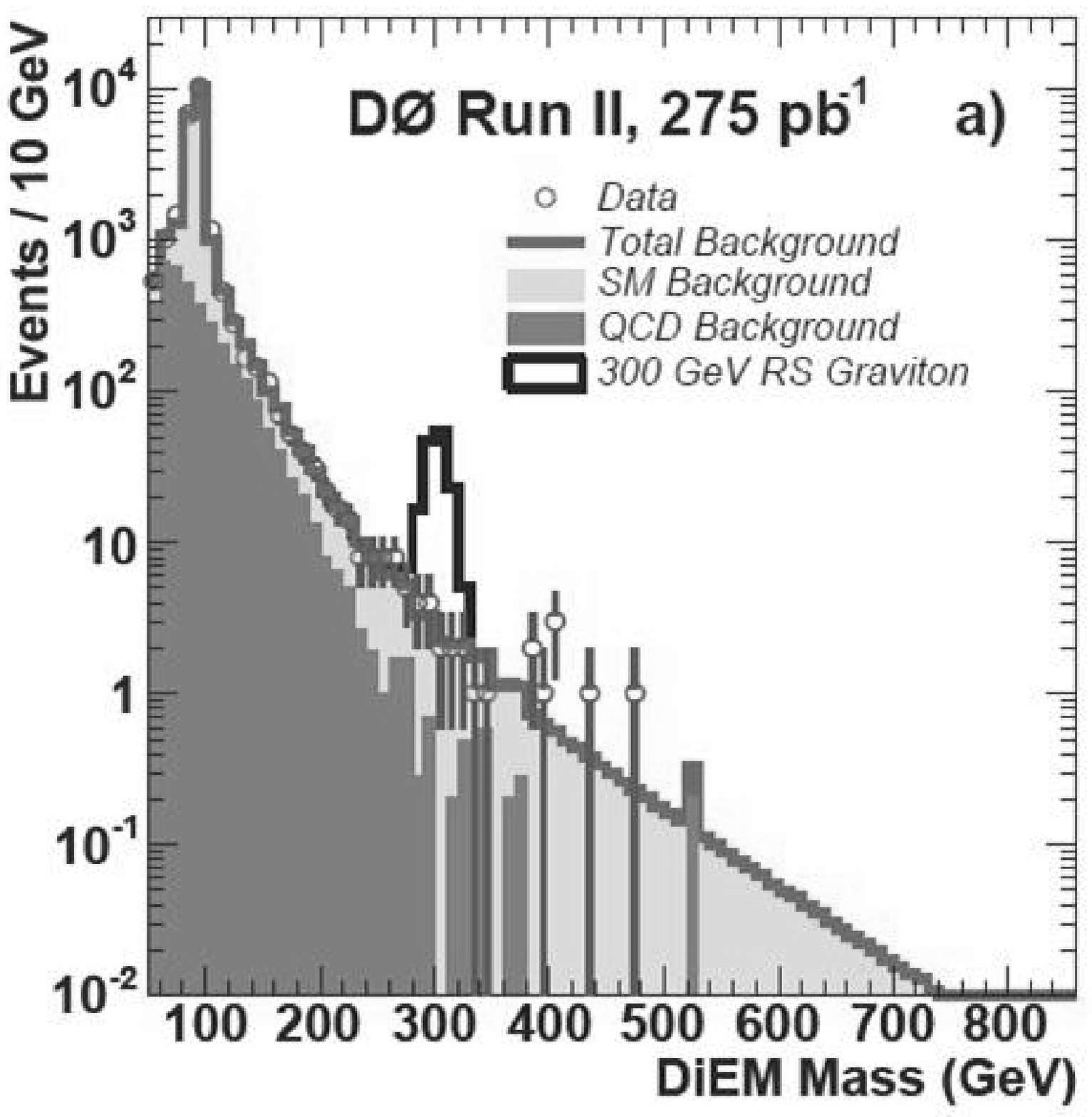}
\end{center}
\begin{center}
\caption{Data/MC comparison (dielectron invariant mass) for the CDF Z' analysis (left) and D0 Randall-Sundrum graviton search (right)}
\label{fig:fig1}  
\end{center}
\end{figure}

Events with two leptons (ee or $\mu\mu$) or two photons may point out to physics beyond the SM. CDF and D0 collaborations performed a search for those final states, heavily relying on the identification of electrons, photons and muons. Electromagnetic objects are requested to be isolated in the calorimeter and the tracker, and have their electromagnetic shower consistent with that expected for an electron or a photon. Isolated muons are identified in the muon spectrometers and are required to have a matching track (quality criteria are applied) in the central tracking detector. The typical identification efficiencies for e,$\gamma$,$\mu$ are of about 80 to \mbox{90 $\%$} for energy ranges considered (\et $>$ 15-20 \gev \hspace{0.05cm} for electromagnetic objects, \pt $>$ 15 \cgev \hspace{0.05cm} for muons). Drell-Yan events ($Z/\gamma^* \rightarrow ee,\gamma\gamma,\mu\mu$) make up the main SM background. For the dielectron/diphoton search, an additional instrumental background due to QCD multijet events and direct photon events has to be carefully estimated from data. Systematic errors are of the order 9 to 13$\%$ for the background (efficiency, muon momentum smearing, PDF). Two kind of studies were performed: a direct search where a resonance (Z', Randall-Sundrum graviton) is produced, and an indirect search where deviations in the dilepton/diphoton production relative to the SM are expected at large invariant mass (Large Extra Dimensions, Compositeness). No excess with respect to the SM expectations has been observed (fig.\ref{fig:fig1}). Limits at the 95$\%$ confidence level were then set for the models considered. For instance a Z' with a mass lower than 850 \ccgev \hspace{0.05cm} is excluded by CDF and a Randall-Sundum graviton with a mass lower than 780 \ccgev \hspace{0.05cm} is excluded by D0, if the graviton coupling to the SM fields is equal to 0.1. Other results are given on the collaboration web pages\cite{refcdf}$^{,}$\cite{refd0}$^{,}$\cite{reftalk}.

\par
CDF has performed a signature-based search looking for $\gamma\gamma+X$ ($X=e,\mu, \gamma$) and $\ell \gamma+X$ ($X$=\met, $\ell$) events. $\gamma \ell+X$ events are requested to contain one isolated photon \mbox{(\et$>$ 25 \gev)} and one isolated central ``tight'' lepton (\et$>$ 25 \gev) plus either \met$>$25 \gev \hspace{0.05cm} or a ``loose'' lepton (\et$>$ 25 \gev). For the $\gamma\gamma +X$ analysis, events with two central photons \mbox{($|\eta|<1$,\et$>$13 \gev)} and either another photon (\et$>$13 \gev) or a lepton (\pt$>$20 \cgev) are selected. High luminosity samples were used for the $\gamma\gamma+X$ analysis: 683 \pb for $\gamma\gamma+e,\mu$ and 1020 \pb for $\gamma\gamma\gamma$. This is the first analysis, in this kind of search, with more than 1 \fb. No evidence was observed as seen on table~\ref{tab:gamma}.

\begin{table}[h]
\caption{Final Results for the CDF signature-based analysis: $\gamma\gamma+X$ and $\ell \gamma+X$\label{tab:gamma}}
\begin{center}
\begin{tabular}{|c|c|c|c|c|c|}
\hline
Final State & $\gamma\gamma\gamma$ & $\gamma\gamma e$ & $\gamma\gamma\mu$ & $\gamma \ell$ \met & $\ell \gamma \ell $\\
\hline
\lumi & 1020 \pb & 683 \pb & 683 \pb & 307 \pb & 307 \pb \\
\hline
Data & 4 & 2 & 0 & 43 & 31 \\
\hline
MC & 1.9 \mypm 0.6 & 4.49 \mypm 0.84 & 0.47\mypm0.12 & 35.1\mypm5.3 & 21.2 \mypm 4 \\
\hline
\end{tabular}
\end{center}
\end{table}

\begin{figure}[htbp]
\begin{center}
\includegraphics[width=4.3cm]{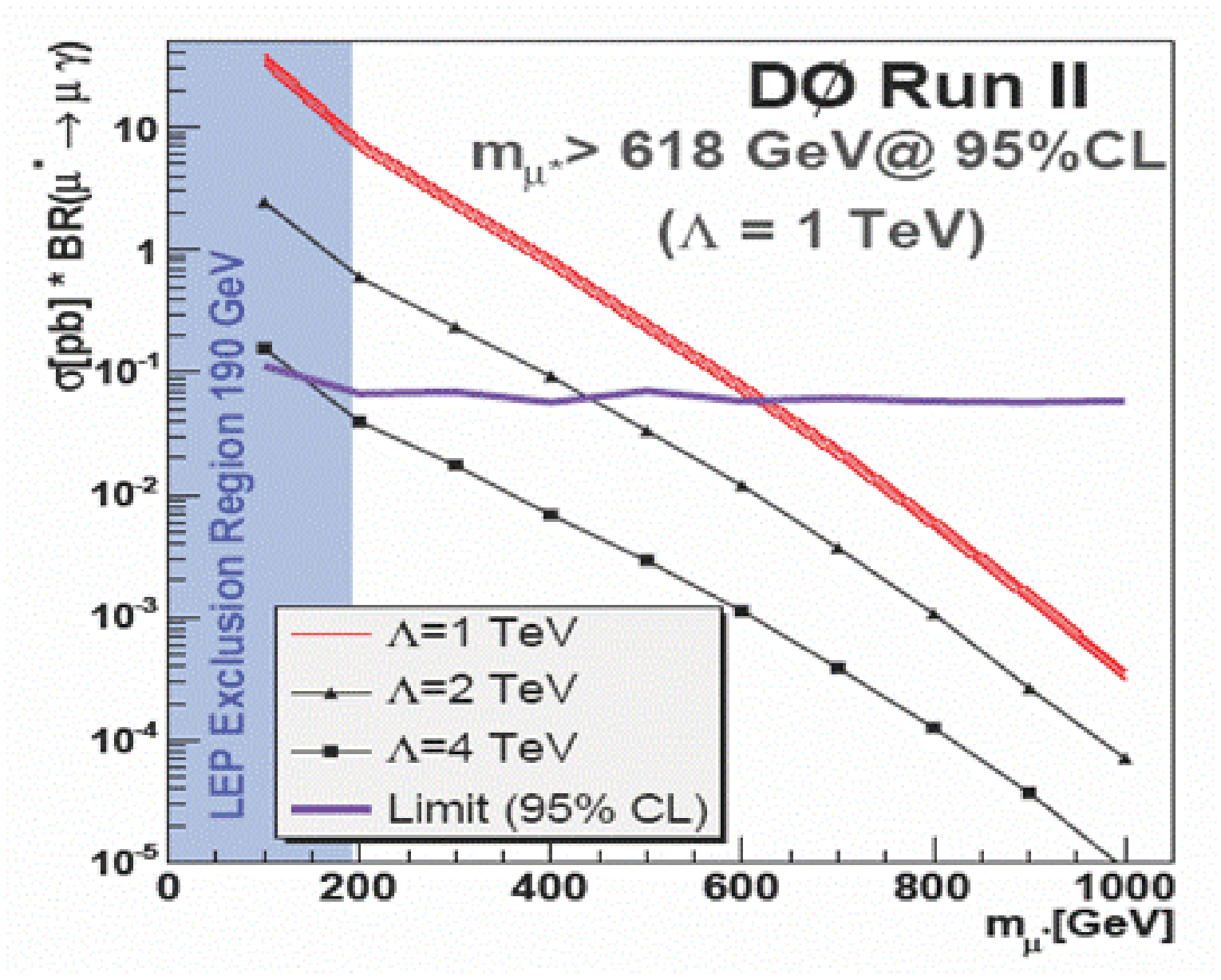}
\includegraphics[width=5.cm]{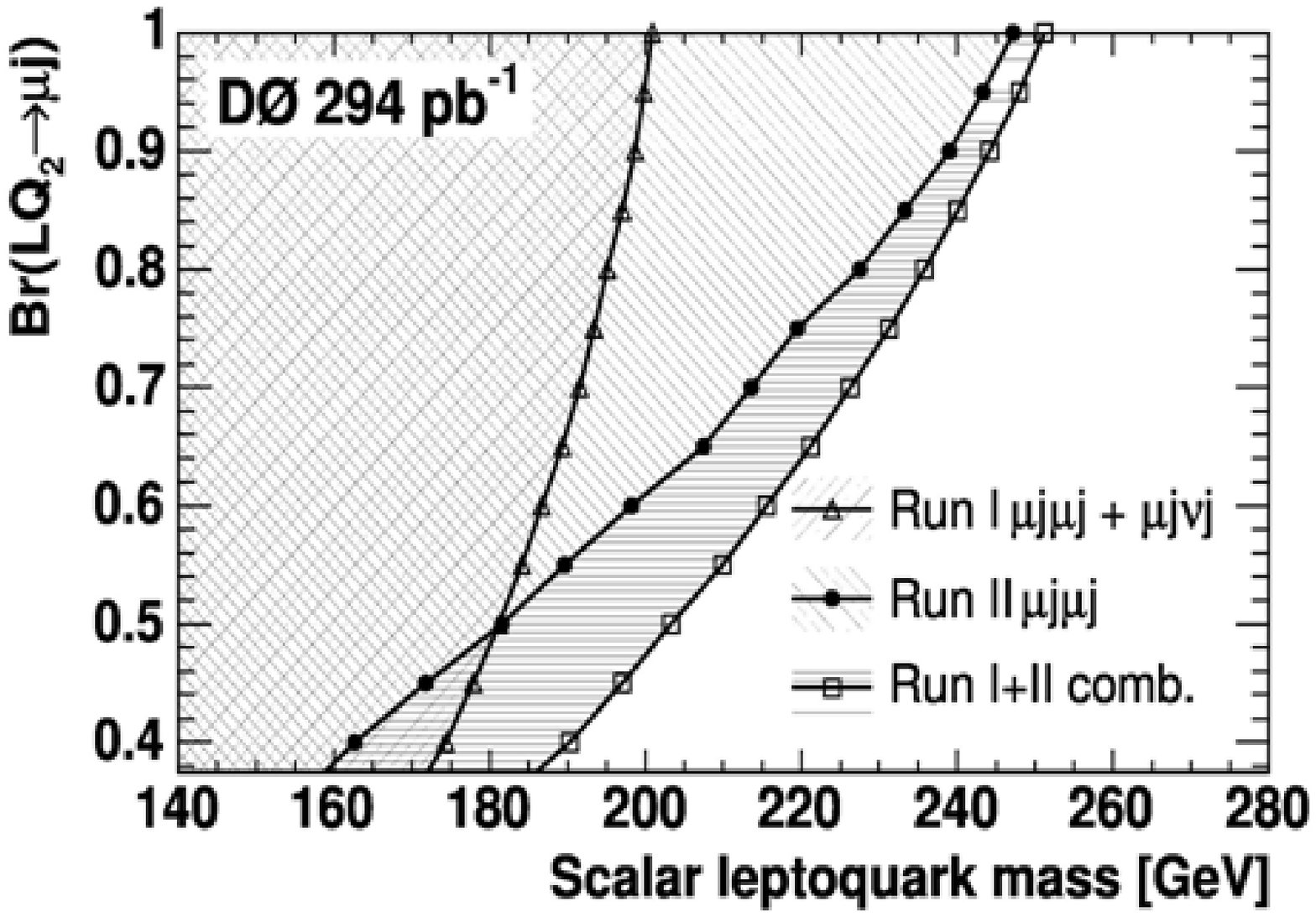}
\end{center}
\begin{center}
\caption{D0 excited muon (left) and second generation leptoquark limits (right)} 
\label{fig:fig2}  
\end{center}
\end{figure}

Excited muons (\mustar) are expected in compositeness models. CDF and D0 collaborations searched for the production \mbox{\ppbar$~\rightarrow\mu$\mustar} with \mustar $\rightarrow \mu \gamma$ (resonant). Two isolated muons (\pt$>$15-20 \cgev) and one isolated photon (\et$>$25-27 \gev) are the first selection cuts of the analysis. The main SM background is made of $Z\gamma, WZ, ZZ$ and $Z+jets$ (fakes) events. To suppress residual background, cuts are applied on the invariant mass of the leading $\mu$ and the photon. No excess was observed. A mass lower than 618 \ccgev \hspace{0.05cm} is excluded by D0 for $\Lambda$=1 TeV (fig.\ref{fig:fig2}) , whereas CDF excludes a mass lower than 800 \ccgev \hspace{0.05cm} for the same compositeness scale. The difference between these two values is to be explained by the way the \mustar decays were considered \cite{refcdf}$^{,}$\cite{refd0}$^{,}$\cite{reftalk}.

\par
Theories based on extensions of the gauge group of the SM, such as left-right symmetric models, predict the existence of additional heavy charged vector bosons as the W'. CDF has performed a search for this particle in the $e\nu$ channel. Events selected contained one isolated electron and missing energy. The SM background is dominated by $W \rightarrow e\nu, W\rightarrow \tau\nu$ and multijets (fakes) processes. The typical efficiency for the signal is 40-45$\%$. No excess was seen, leading to an excluded mass lower than 788 \ccgev \hspace{0.05cm} at 95$\%$ confidence level.

\section{Leptons+jets final state}

Leptoquarks (LQ) are bosons carrying the quantum numbers of a quark-lepton system. Searches for second generation Leptoquarks in the channel $LQ_2 LQ_2 \rightarrow \mu\mu jj$ (CDF+D0) and $\mu\nu jj$ (CDF) were performed. The final state of interest is made of two isolated energetic jets (\et$>$15 to 30 \gev) and one or two isolated muons (\pt$>$15 to 25 \cgev). A cut on the $\mu\mu$ invariant mass is done for the $\mu\mu jj$ analysis and \met$>$60 \gev \hspace{0.05cm} is requested for the $\mu\nu jj$ channel. The residual background (Drell-Yan: $Z/\gamma^* \rightarrow \mu\mu+jets$) vanishes with additional topological cuts. No excess with respect to the SM expectations was observed. A second generation Leptoquark mass of 251 \ccgev \hspace{0.05cm} was excluded for $\beta$=1 ($\beta$ is the branching ratio of the decay $LQ_2\rightarrow l^{\pm}q$)(fig.\ref{fig:fig2}).

\par
Production and decay of excited quarks may also lead to the leptons+jets final state. D0 looked for the process $gq\rightarrow q^{*} \rightarrow Zq \rightarrow eeq$. Events with two electrons (\et$>$30, 25 \gev) and one jet (\pt$>$10 \cgev) were selected. A cut on the dielectron invariant mass was performed to remove the SM dominant background (Drell-Yan to $ee$). No excess was observed. An excited quark mass lower than 520 \ccgev \hspace{0.05cm} was excluded.

\section{Jets+\met final state}
Jets come from final state partons that are produced in hard \ppbar collision. Few effects produce a distortion in the jet energy when compared to its particle level source: jets are made up of different kind of particles for which calorimeter response are different and there are also instrumental (electronic noise) and reconstruction (jet algorithm) contributions. All these effects are part of the Jet Energy Scale (JES), which must be fully understood to perform analysis with jets+\met in the final state.
\par
\begin{figure}[htbp]
\begin{center}
\includegraphics[width=5.9cm]{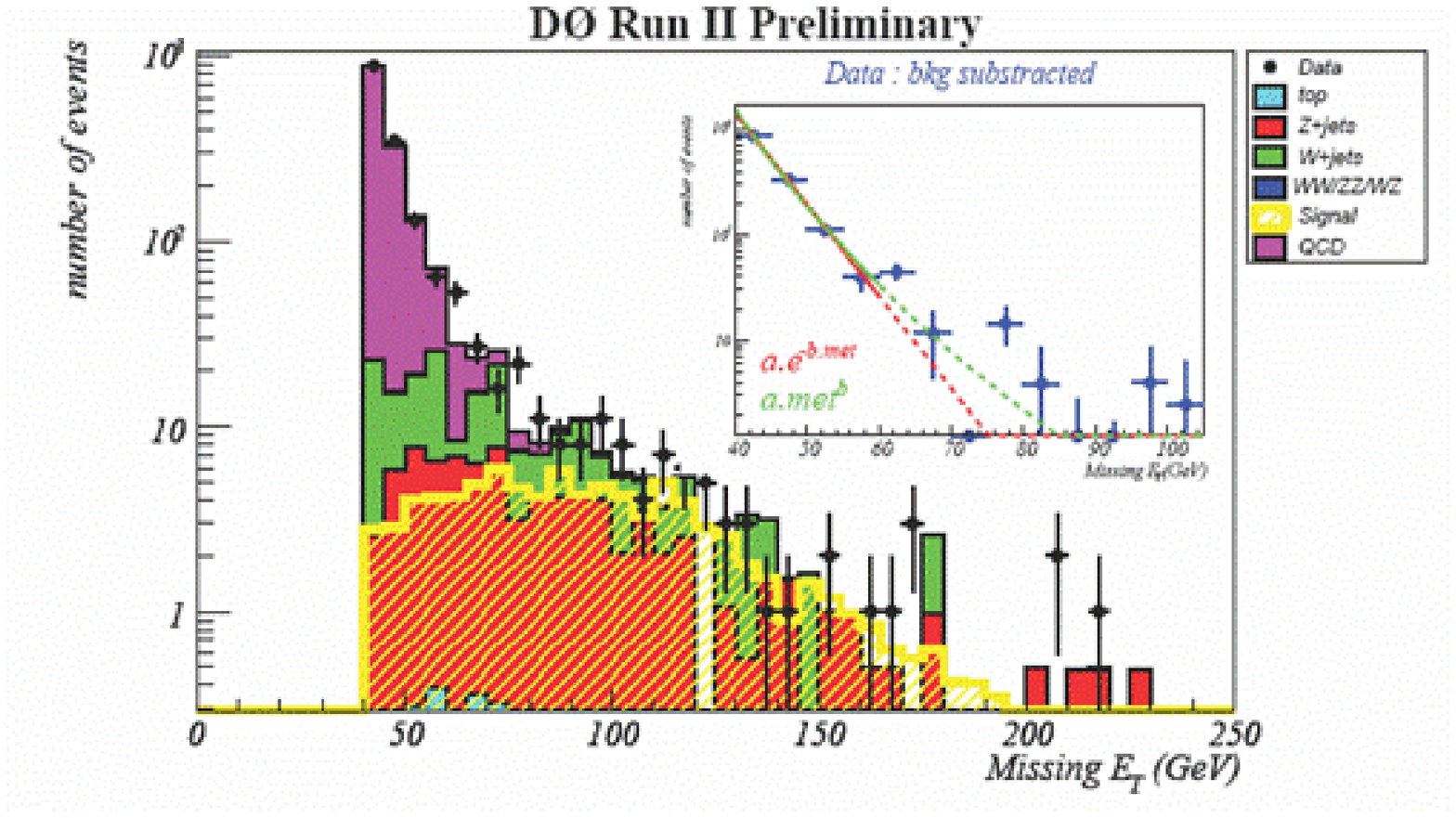}
\includegraphics[width=4.5cm]{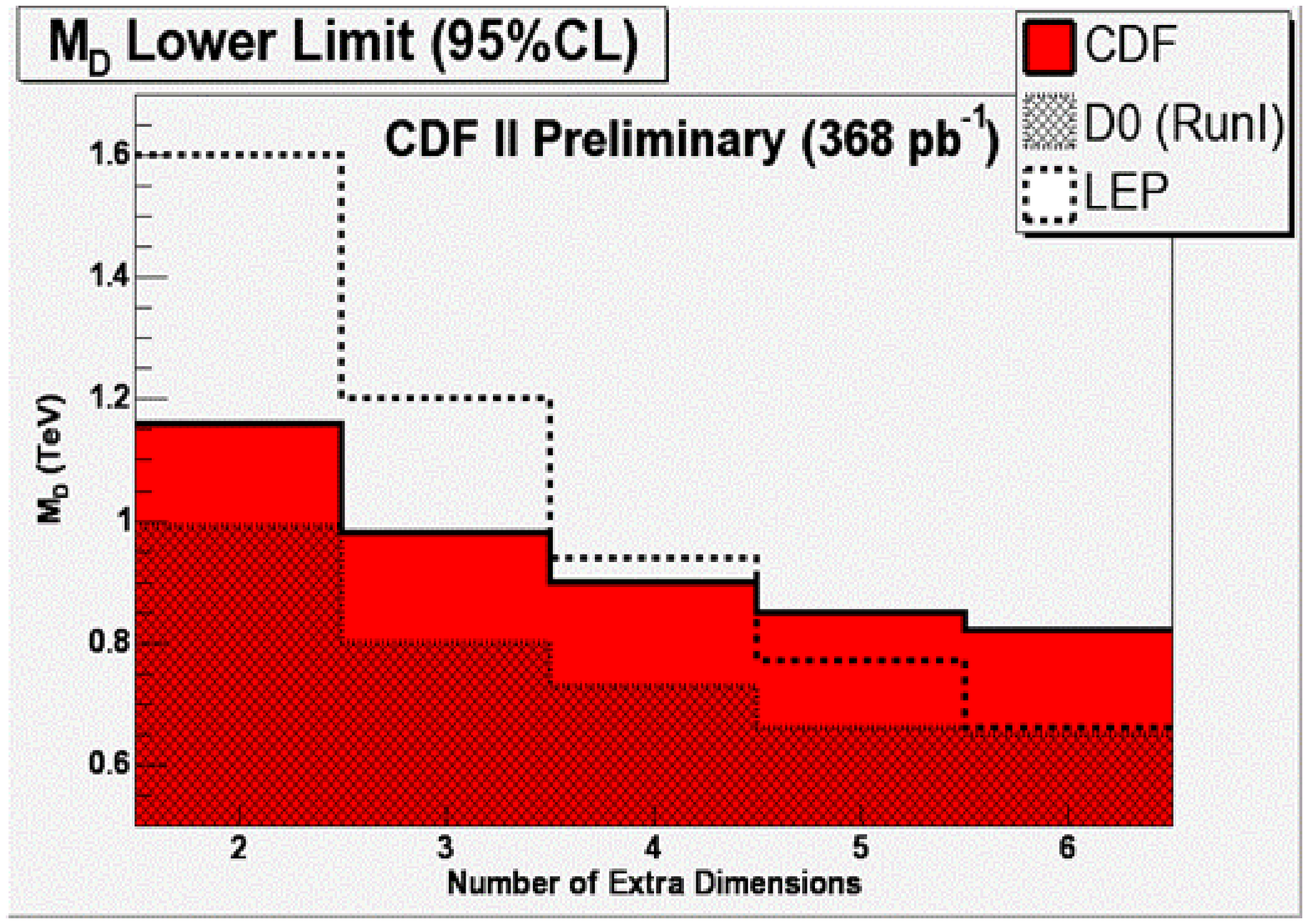}
\end{center}
\begin{center}
\caption{\met distribution for the D0 scalar leptoquark analysis (left) and Large Extra Dimensions CDF limits (right)} 
\label{fig:fig3} 
\end{center}
\end{figure}

  
CDF and D0 collaborations performed a search for scalar leptoquarks through the process \mbox{\ppbar $\rightarrow LQLQ \rightarrow \nu\nu jj$} giving a signature of two acoplanar light jets and missing energy. Events with two jets of fairly high \pt (40 and 25 \cgev \hspace{0.05cm} for CDF, 60 and 50 \cgev \hspace{0.05cm} for D0) and large \met (60 \gev \hspace{0.05cm} for CDF and 80 \gev \hspace{0.05cm} for D0) were selected. The main background, $Z(\nu\nu)+jets$,$W(l\nu)+jets$ and QCD multijets (instrumental), was removed by additional suitable cuts. A good agreement between data and MC was observed (fig.\ref{fig:fig3}). A mass lower than 136 \ccgev (117 \ccgev) for $\beta$=0 was excluded by D0 (CDF). The systematic errors are of the order of 15$\%$ and are dominated by the luminosity measurement, the JES, the jet energy resolution, and the PDF.
\par
Events with jets and \met are predicted by Large Extra Dimensions models, when (g,G) or (q,G) pairs are produced: the graviton (G) propagates in extra dimensions, leading to a signature of a single energetic jet and \met. To minimize the sensitivity to the JES, CDF has estimated the main background ($Z(\nu\nu)+jets$,$W(l\nu)+jets$ and QCD multijets) from data. Events were then selected by requesting a very high \pt jet (150 \cgev) and very large \met (150 \gev). Additional cuts were applied to remove the background. A fairly good agreement between data and MC was observed: 263 events observed while 265$\pm$30 were expected. The total uncertainty is quite low (11$\%$ for statistic and systematic) because of the way the background was estimated. Limits depending on the parameters of the model were drawn (fig.\ref{fig:fig3}).

\section{Summary}

A comprehensive set of non-supersymmetric searches is performed at the TeVatron. The signatures studied, leptons/photons, leptons+jets and jets+\met, request a good understanding of the detector to keep systematic uncertainties at low levels. No excess with respect to the Standard Model expectations has been observed and parameter regions have been excluded. Most of the results are world best limits (Extra Dimensions, excited muons, W', second generation leptoquarks, scalar leptoquarks). CDF and D0 performed an impressive set of research and all the topics could not be detailed here. Further information can be found on the public web pages of the two experiments \cite{refcdf}$^{,}$\cite{refd0}. High luminosity samples corresponding to ten times the RunI luminosity (namely 1 \fb) are being analyzed and new exciting results are in the pipeline. 

\section*{Acknowledgments}
I thank my colleagues from CDF and D0 collaborations for providing the results. I also thank the organizers of the conference for their hospitality.

\end{document}